\documentclass[12pt]{article}
\pdfoutput=1
\usepackage{setspace}
\usepackage{color}
\usepackage{amsmath, amssymb,slashed,url}
\usepackage{amsthm}
\usepackage{graphicx,epsfig}
\usepackage{cite}
\usepackage{jheppub}
\usepackage{color}

\newcommand{\beq}{\begin{equation}}
\newcommand{\eeq}[1]{\label{#1}\end{equation}}
\newcommand{\bea}{\begin{eqnarray}}
\newcommand{\eea}[1]{\label{#1}\end{eqnarray}}

\def\draftnote #1{{\color{red} #1}}

\begin{document}

\rightline{{Imperial-TP-2018-CH-01}}

\vspace{16pt}

\begin{center}

{\huge \bf Superstrings on AdS$_3$ at $k=1$}

\vskip 20pt

{\large   
G. Giribet$^a$,
C. Hull$^b$, M. Kleban$^a$, M. Porrati$^a$ and E. Rabinovici$^{c,d}$}

\vspace{12pt}

{$^a$\em Center for Cosmology and Particle Physics, \\ Department of Physics, New York University, \\726 Broadway, 
New York, NY 10003, USA}\footnote{email: gg1043@nyu.edu, mk161@nyu.edu, massimo.porrati@nyu.edu}

\vspace{12pt}

{$^b$\em The Blackett Laboratory, Imperial College London \\
Prince Consort Road London SW7 2AZ, U.K.}\footnote{email: c.hull@imperial.ac.uk}

\vspace{12pt}

{$^c$\em Racah Institute of Physics, The Hebrew University of Jerusalem, 91904, Israel}\\ $^d${\em IHES, Bures-sur-Yvette, France}\footnote{email: eliezer@mail.huji.ac.il}

\end{center}

\begin{abstract} 
 
 \noindent 
We study superstring theory in three dimensional Anti-de Sitter spacetime with NS-NS flux, 
focusing on the case where the radius of curvature is equal to the string length. 
This corresponds to the critical level $k=1$ in the {formulation as a Wess-Zumino-Witten model}.  Previously, 
it was argued that a transition takes place at this special radius, from a phase dominated by black holes at
 larger  radius to one dominated by long strings at smaller radius.
We argue that the infinite tower of modes that become massless at $k = 1$ is  a signal
of this transition. We propose a simple two-dimensional conformal field theory as the
holographic dual to superstring theory at $k=1$. As evidence for our conjecture, we
demonstrate that our putative dual exactly reproduces the full spectrum
of {the long strings of the weakly coupled string theory}, including states unprotected by supersymmetry.

\end{abstract}

\section{Introduction}\label{intro}

String theory in three-dimensional anti-de Sitter spacetime AdS$_3$ 
has a dimensionless coupling $k = R_{AdS}^2/l_s^2$, where $R_{AdS}$ is the radius of curvature of AdS$_3$ and $l_s$ is the string length.  The theory is well-understood in the semi-classical regime of large $k$.
Our purpose here is to explore the 
interesting phenomena that occur in the 
stringy regime at a specific small value of $k$
where $R_{AdS} \sim l_s $.
It was  argued in  \cite{gkrs} that there is a critical value $k = k_c$ (given by $k_c=3$ for the bosonic string and $k_c=1$ for the superstring with NS-NS flux) at which
the theory undergoes a transition from a phase  $k>k_c$ dominated by black holes to a phase $k<k_c$ dominated by long strings. This was based on both UV and IR considerations.
At this same value $k=k_c$ an infinite tower of higher spin states become massless \cite{Gaberdiel:2017oqg}.
This suggests that the appearance of the massless higher spin states is the IR manifestation of a stringy phase transition.  
 {The linear dilaton background of 
\cite{gkrs} has a similar phase transition, and also appears to have an infinite tower of massless states at $k=k_c$.} 

A phase of string theory in which there are an infinite number of massless higher spin states would be a highly symmetric phase in which there are unbroken stringy symmetries and which could reveal much about the fundamental structure of string theory. It has been argued that such phases could be probed in high-energy large-angle scattering \cite{Amati:1987wq,Gross:1987ar} or for strings in anti-de Sitter space in a limit where the radius of curvature is {much smaller than  the string length \cite{Sundborg:2000wp}, \cite{Witten}, \cite{Mikhailov:2002bp}, or of order of the string length \cite{Lindstrom:2003mg}}. 
 In both these cases, the massless states can be associated with fundamental strings becoming tensionless.
However for strings in AdS$_3$ at  $k=k_c$ the situation is rather different: {it is long strings near the boundary that become tensionless.}

 {Some states in the short-string spectrum, including the $L_2$-normalizable vacuum and the graviton, are absent from the spectrum when $k \leq k_c$.  Precisely at $k=k_c$,  the short-string vacuum and graviton are replaced by long-string states with the same spacetime quantum numbers, along with an infinite tower of massless higher spin states.  As long strings, all these states are plane-wave rather than strictly normalizable, and lie at the bottom of a continuum of states 
 parameterised by a continuous variable $\sigma$.
 (For large $k$, this variable
  can be interpreted as the radial momentum of long strings.) 
  It was
 argued in \cite{gkrs} that for $k< k_c$ the spectrum does not contain either black holes or other non-perturbative
 excitations, so that the phase with $k < k_c$ would be a novel theory 
 dominated by the long-string degrees of freedom. }
 
 The special features of the spectrum at $k=k_c$ provide 
 clues 
 as to the nature of the holographic dual conformal field theory
  (CFT).  In this paper we first discuss the physics of the theory near $k=k_c$, and then identify a CFT that 
  exactly reproduces the \emph{full} 
spectrum (not just the BPS spectrum)   of the long strings of {the} weakly-coupled string theory in AdS$_3$ at this critical level.

This paper is organized as follows: Section 2 describes the phase structure of string theory on a 
WZW $AdS_3$ space. The
phase transition at $k=k_c$ is discussed in some detail. Section 3 studies in greater detail the  spectrum of the
$N=1$ superstring at $k=k_c=1$. Section 4 proposes a 2-d CFT dual
to the $k=1$ $AdS_3$ superstring. It is a symmetric-product orbifold, whose complete 
spectrum we describe exactly. Section 5
compares the spectrum of the   long strings on $AdS_3$ with the spectrum of the proposed holographic dual CFT and finds
perfect agreement. Section 6 presents  a preliminary comparison of interactions in the two theories. A potential 
disagreement is shown to exist and possible 
 resolutions 
are discussed.   
In Section 7 we give some concluding remarks.

\section{The Phase Structure of String theory in $AdS_3$.}

A useful key to unravelling the phase structure and high energy behavior of a system is to determine which sets of degrees of freedom  control the dynamics at different scales.
A given system will typically have many kinds of degrees of freedom -- e.g. particles, solitons, radiation, black holes, strings etc --  and the contribution to the  entropy associated  with each will depend on the temperature or energy, coupling constants etc. 
In a given regime, there will typically be one set of degrees of freedom that dominates the entropy. There will then be various  regimes in each of which  a different degree of freedom dominates, corresponding to different phases of the theory.
  
In gravity, the key players are light particles, radiation, strings, branes and  various black objects. Each player indeed dominates for some energy range and this realization has led to  interesting phase diagrams for gravitational systems. 
A particularly interesting transition is that between  phases dominated by black holes and phases dominated by strings. For dimensions $D>3$, the intuitive arguments  involved in describing the transition  relate to the change in the curvature at the black hole
horizon as a function of the energy/mass of the configuration: the weaker the horizon curvature, the more the black hole description is justified. 
The transition between the black hole picture,  which is  valid for small horizon curvature, and the perturbative string picture, which is valid for large curvature,  occurs when the curvature at the horizon is of order the string scale and is smooth.

For $D>3$, the horizon curvature    of a black hole depends on its mass, and goes to zero for large mass.
This  does not apply for black holes in AdS$_3$ for which
the curvature depends only on the cosmological constant -- it is the same everywhere in the spacetime and is  independent of the black hole mass.
Accordingly, a  somewhat different approach was   proposed in~\cite{gkrs} for the case of string theory on an AdS$_3$  background.
The focus was  on the   curvature scalar,  which is a constant given by the cosmological constant 
$$\Lambda=-{1\over R_{AdS}^2} \, = -{1\over k l_s^2}.$$ 
For $D=3$,  one can then use  $k$  as a tuneable parameter  to probe
  the string/black hole transition.  
 
 In AdS$_3$, there is an asymptotic space-time Virasoro algebra with central charge  $c$, given for large $R_{AdS}$ by 
 \beq
 c={3R_{AdS}\over 2l_p},
 \eeq{cent}
 where $l_p$ is the three dimensional Planck length.
 The entropy  behaves at
high energies as
\beq
S=2\pi\sqrt{c_{\rm eff}L_0\over6}+
2\pi\sqrt{\bar c_{\rm eff}\bar L_0\over6},
\eeq{entis}
for some
 effective central charges $c_{\rm eff}, \bar c_{\rm eff}$,
where $L_0$ and $\bar L_0$ are the
asymptotic Virasoro generators related to the energy $E$
and angular momentum $s$ in AdS$_3$ by
 \beq
 ER_{AdS}=L_0+\bar L_0-{c\over12};\;\;\;s=L_0-\bar L_0.
 \eeq{enis}
 Unitarity implies that the effective central charge $c_{\rm eff}$ satisfies
 \beq
 c_{\rm eff}\le c
 \eeq{erw}
 and $c_{\rm eff}=c$
if and only if the $SL(2,\mathbb{C})$ invariant vacuum is normalizable.  
 
In general, the spectrum of strings in AdS$_3$ includes excited short strings and long strings near the boundary. In addition there are BTZ black hole states and possible further states  such as branes wrapping additional dimensions that depend on the details of the string background AdS$_3\times {\cal N}$~\cite{Kleban:2013wba}. 
 
 In~\cite{gkrs}, several pieces of evidence were discovered that pointed to the value of $k=k_c$ as the string/black hole transition point.  The first is related to the proper identification of the effective degrees of freedom.
{For $k<k_c$ the dominant degrees of freedom at all energy scales are the long strings. There is no graviton and no states corresponding to BTZ black holes.} In this regime, $ c_{\rm eff}<c $ and the
$ SL(2,\mathbb{C}) $ invariant vacuum is not normalizable, and so  perturbative string states are also not normalizable.
In particular, there is not a normalizable graviton state.
For $k>k_c$, $ c_{\rm eff} =c$ and the $ SL(2,\mathbb{C}) $ vacuum and perturbative string states
are normalizable. The BTZ black hole states are  also normalizable and  dominate the entropy at high energy~\cite{gkrs},\cite{Kleban:2013wba}.

\begin{figure}
  \centering
  \includegraphics[width=.5\textwidth]{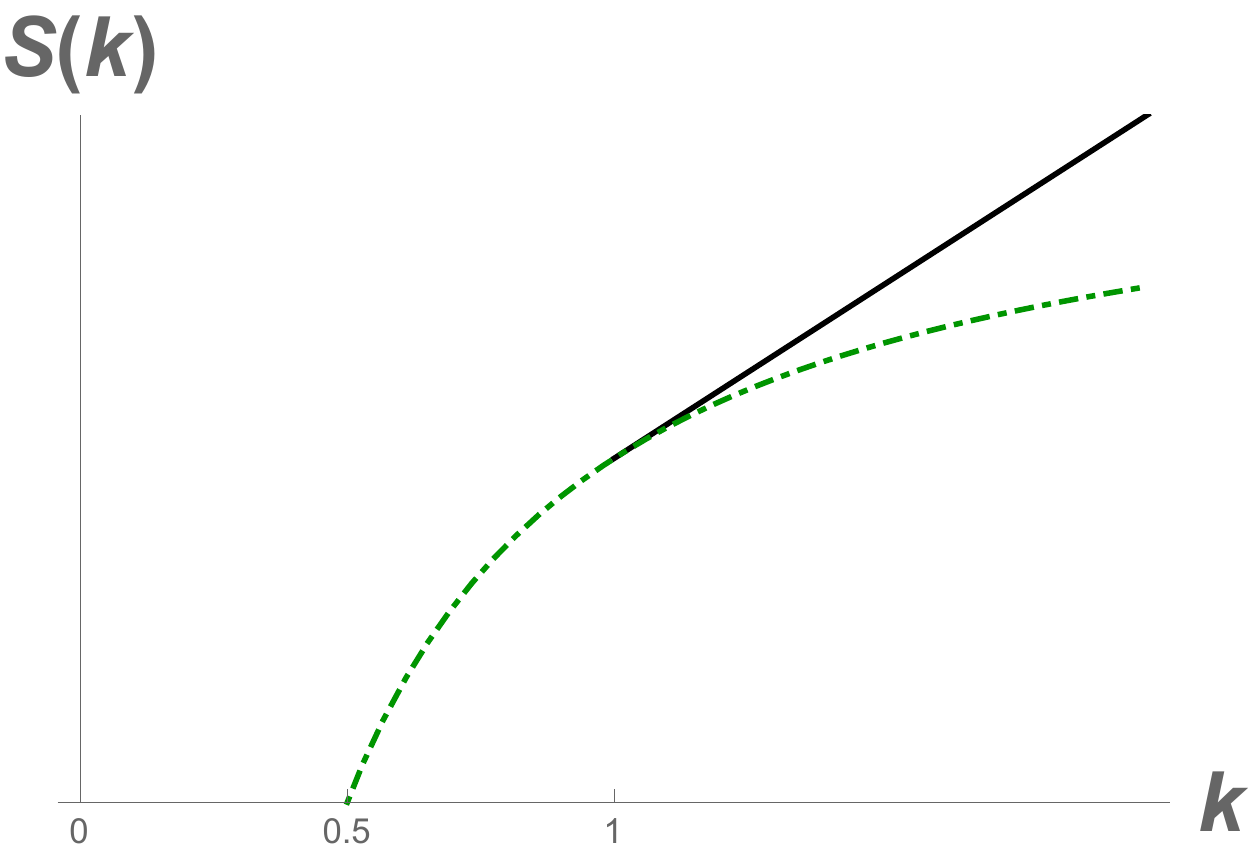}
  \caption{\small The behavior of the entropy $S$ at high energies, as a function of the WZW level $k$.    The black solid line is the entropy of black holes, which do not exist for $k<1$.  The dashed line is the entropy of long strings, that dominate  for $k<1$ (see \cite{gkrs} for details).
  \label{figure}}
\end{figure}

The entropy of the black  holes for $k>k_c$ and of the long strings for $k<k_c$ matches exactly at $k=k_c$, but the first derivative of the entropy with respect to $k$ is not continuous.
The fact that  $ c_{\rm eff} =c$  for $k>k_c$, $ c_{\rm eff}<c $ for $k<k_c$, and $ c_{\rm eff}$ is continuous  through the critical value 
 suggests that
$ c_{\rm eff} $ can be considered  an order parameter for this transition. 
Finally,  the effective coupling of the long strings near the boundary switches from
being weak if $k<k_c$  to becoming strong if $k>k_c$.  The entropy matching is a UV property, and the behavior of $c_{\rm eff}$ can be considered to be  an IR 
property.

Thus there is some evidence for a  transition at $k=k_c$ from a black hole phase to a phase dominated by long strings. Note that this is different from the usual string/black hole transition in which the  string phase is dominated by highly excited short strings that for large excitation number begin to resemble a black hole~\cite{Horowitz:1996nw, Barbon:2004dd}.  Here the  strings are   of  a different nature, long and localized near the boundary of AdS$_3$~ \cite{mo}.
For further confirmation
 of the presence of a string/black hole phase transition, one would want to identify the IR signature of a phase transition, for example  the emergence of massless excitations at the transition point. 
 Such massless modes were
discovered in~\cite{Gaberdiel:2017oqg} and we 
argue that these are indeed the signal of a phase transition. 

 Before discussing these modes in more detail, it will be useful to discuss the differences between second order phase transitions in   QFT and those  in  string theory.  In a QFT one typically  varies a  coupling to a  relevant operator to go from one phase to another, passing through a point or surface  in parameter space in which massless degrees of freedom play a role (in some cases they play a role in a whole region of parameter space). By contrast, perturbative string theory  is heavily constrained -- the worldsheet theory must be a quantum conformal field theory  with specific features, among them a specific value for the central charge.
  {The worldsheet theory of strings on AdS$_3$}
must be accompanied by a unitary CFT, so that as one varies the value of the level $k$ one must also vary the attendant CFT so as to keep the total central charge unchanged  {at the critical value}. 
In the QFT case one can have Goldstone modes or a 
  {Brout-Englert-Higgs mechanism}  
at and around the fixed point with a finite number of massless particles involved. 
It is not a priori clear exactly what to  expect for a transition in a string theory. The results obtained 
in~\cite{Gaberdiel:2017oqg}  for the bosonic case and in~\cite{Gaberdiel:2017oqg,gaberdiel-ads} for the supersymmetric case
give  two chiral trajectories of a discrete infinity  of massless states with an ever growing spin at the  same transition point $k=k_c$.
This is the desired IR manifestation of the  transition.

\section{String Theory in AdS$_3$}

The bosonic string in AdS$_3$ is formulated in terms of a WZW model with target space  AdS$_3$ (the universal covering space of the group manifold  $SL(2,\mathbb{R})$) with level $k$.  The complete  perturbative spectrum (at weak string coupling but to all orders in the string tension $\alpha'$) was found in \cite{mo}.
  The superstring in AdS$_3$ with NS-NS flux is given by a supersymmetric WZW model, and
   has been analyzed in
    \cite{ gaberdiel-ads, Giribet:2007wp, Giveon:1998ns,Pakman:2003cu,Israel:2003ry,Raju:2007uj,Argurio:2000tb}.
    The bosonic string and the superstring with NS-NS flux in AdS$_3$ are formally very similar, but the bosonic string has  a 
  tachyon  \cite{mo}  which leads to an exponential divergence in the partition function and presumably renders the theory unstable.
In both cases, the spectrum consists of both discrete and continuous representations of   $SL(2,\mathbb{R})$, and a ``spectral flow" acts on the representations.
The excitations of the fundamental strings fit into discrete representations, but there are also long strings near the boundary of AdS$_3$ that are associated with spectrally-flowed continuous representations \cite{mo}.
We shall first review the bosonic string, which introduces many of the key features, and then turn to the superstring.

\subsection{The  Bosonic String on $AdS_3$.}

The Virasoro central charge for the $ {SL}(2,\mathbb{R})$ WZW model at level $k$ is 
\beq
c_k = \frac{3\, k}{k-2} \ . 
\eeq{sdfa}
For a critical string theory, we must consider the tensor product of this with a CFT $\cal{N}$ with central charge $c_N= 26- c_k$.
For a unitary theory  $c_N\ge 0$, which implies
\beq
k\ge 52/23.
\eeq{sdfgh}
 This in particular excludes $k=2$, so that the $k=2$ $ {SL}(2,\mathbb{R})$  WZW model 
cannot be part of a conventional critical string theory.
As mentioned previously, if we are to consider the behavior of the theory as $k$ changes,  
then we have to change the CFT $\cal{N}$ as we vary $k$,  so that the central charge of $\cal{N}$
remains $c_N= 26- c_k$. 
For $k=3$, we note that $c_k=9$ and $c_N=17$.

States are labelled by  the eigenvalue $C$  of the quadratic Casimir   of $ {SL}(2,\mathbb{R})$  and by the eigenvalue $m$ of of $J^3_0$ for the left-handed $ {SL}(2,\mathbb{R})$, together with the corresponding quantities $\bar {C}$ and $\bar m$ for the right-handed $ {SL}(2,\mathbb{R})$.
They are also labelled by an integer $w$ characterising the spectral flow; $w$ is associated with the winding number for long strings \cite{mo}.
The spacetime energy $E$ and spin $s$ are
\beq
E = m+\bar m \ , \qquad s = m- \bar m \ .
\eeq{sdfgghj}
The AdS$_3$ 
mass is given by: 
\beq
m^2_{{\rm AdS}_3} = (E - |s|) ( E + |s| -2) \ . 
\eeq{dict}
States for which $E=|s|$ are then massless higher spin fields.

For $w=0$, there are two kinds of representations that appear in the spectrum.
First, there are the discrete lowest weight representations ${\cal D}^+_j$ for which 
$m= j, j+1, j+2, \ldots$ and $C =
 -j (j-1)$ (with $j$ {a positive real number}).  These correspond to the so-called short string states.
Secondly, there are the continuous 
representations ${\cal C}(\sigma ,\alpha)$ labelled by
 $\sigma\in\mathbb{R}$, $\sigma> 0$, $\alpha\in\mathbb{R}$ for which  $m$ takes the values
$m\in\alpha+\mathbb{Z}$  and $C =  \frac{1}{4} + \sigma^2$. These correspond to the long string states.  Without loss of generality, the 
real parameter $\alpha$ can be taken to be in the range $0\le \alpha <1$. Spectral flow then generates versions of these representations with $w\ne 0$.  
The limit $\sigma \to 0$ gives a reducible representation that can be
decomposed into the direct sum of highest- and lowest-weight discrete representations. {All string states are in these two classes of representation, and  a unitary string theory with a modular invariant partition function is obtained by imposing the Virasoro constraints.

The mass-shell condition for a state of given $C,m,w$ and excitation number $N$ is  \cite{Gaberdiel:2017oqg, mo}
\beq
\frac{C } {k-2}  - w \tilde m -\frac{k}{4} w^2 +h_{\rm int} +N = 1 \ , 
\eeq{massflow}
where $h_{\rm int}$ is the eigenvalue of the $L_0$ operator for the internal CFT  ${\cal N}$
and $\tilde m$ is the eigenvalue of the spectrally flowed $J_0$ operator, related to $m$ by
\beq
m= \tilde m + \frac k 2   w
\eeq{flowedm}
A similar equation arises for the right-moving operators.

{Reference~\cite{mo} shows that the constraints from requiring unitarity,  normalizability of physical states, and symmetry under spectral flow
imply that short string physical states exist only for 
\beq
1/2<j<(k-1)/2.
\eeq{unitarity}
The analysis of the partition function in \cite{mo2} supports this conclusion. For $k\leq 3$ this excludes many short string states, including the vacuum and the graviton with $j=1$.  For $k=3$, these are replaced by plane-wave normalizable  long string states with the same spacetime quantum numbers.

The massless higher spin states identified in \cite{Gaberdiel:2017oqg}   arise from the continuous representation with  }
 \beq
C =\bar C=  \frac{1}{4} + \sigma^2, \qquad
w=1, \qquad   h_{\rm int}=\bar  h_{\rm int}=0
\eeq{cons}
Then the physical state conditions give
\beq
m= N+\sigma^2, \qquad \bar m= \bar N+\sigma^2.
\eeq{masssh}
The continuous parameter $\sigma$ is real and positive;
massless states then arise if $m=0$ or $\bar m=0$, i.e. if $\sigma^2=0$ and either $N$ or $\bar N=0$.
If $\bar N=0, \sigma^2=0$, these have $E=s=N$ and there is a massless higher spin state of spin $N$ for each   positive integer $N$. These arise as part of a continuum of states labelled by $\sigma$ with
$E=N+2\sigma^2$ and spin $s=N$ which are massive if $\sigma\ne 0$.
Similarly, if $N=0$, there is a continuum of  states with $E=\bar N+2\sigma^2$ and spin $s=- \bar N$ which are massless for $\sigma=0$.

It was suggested in
 \cite{Gaberdiel:2017oqg}  that the  stringy tower of  higher spin states that become massless in the limit $k\to 3$ might correspond to a symmetry  structure similar to that of the Higher Spin Square 
\cite{Gaberdiel:2015mra,Gaberdiel:2015wpo}.  

This fits  with the  picture of Seiberg and Witten
\cite{Seiberg:1999xz}, who
  argued that  a long fundamental string in AdS$_3$ is effectively described, for 
   large $k$,
   by a Liouville
theory with background charge $Q$ and central charge $c=1+3Q^2$, where
\beq 
Q= (k-3) \sqrt{ \frac 2 {k-2}} \ . 
\eeq{sdgjjo} 
The Liouville field corresponds to the distance of the long string from the boundary of AdS$_3$.
It  has a spectrum that  is continuous above 
 a threshold energy 
\beq
\Delta_0=\frac{Q^2}{8} =\frac{(k-3)^2}{4(k-2)} \ .
\eeq{liou}
In the limit $k\to k_c=3$ 
this threshold energy goes to zero, giving massless states as part of a continuum, so that the effective theory  
reproduces some of the features of the exact analysis. 
In this limit, a {pure}  Liouville description becomes problematic, as the Liouville theory is likely no longer a unitary CFT for $Q<2\sqrt2$. {Unitarity may be recovered when the Liouville theory is a sector of a larger CFT
 which  contains other marginal operators besides the Liouville potential~\cite{ks}. Moreover, the central charge and the 
 threshold energy are robust properties that hold beyond the large-$k$ limit~\cite{Seiberg:1999xz}; therefore, we} expect that the effective CFT for the long string should be a $c=1$ CFT in the limit $k\to k_c=3$.\footnote{Early work that highlighted the interesting phenomena in the WZW model at $k=3$ and the resemblance to Liouville coupled to $c=1$ matter is \cite{Giribet:2004zd}.}

\subsection{Superstrings on AdS$_{3}$} \label{sss}

The situation for the superstring is similar   \cite{ gaberdiel-ads, Giribet:2007wp, Giveon:1998ns,Pakman:2003cu,Israel:2003ry,Raju:2007uj,Argurio:2000tb,Hoare:2013ida,Lloyd:2014bsa,Baggio:2017kza}. The  $ {SL}(2,\mathbb{R})$ $N=1$ supersymmetric WZW model at level $k$ has
 super-Virasoro central charge 
\beq
c_k = \frac{9}{2}  + \frac{6}{k} \ . 
\eeq{ssdfa}
For a critical string theory this must be tensored with a SCFT $\cal{N}$ with central charge $c_N= 15- c_k$.  For a unitary theory with $c_N\ge 0$, we need
\beq
k\ge \frac 4 7.
\eeq{sdfghfj}
For $k=1$, $c_k=21/2$ so that $c_N=9/2$. For instance, $\cal{N}$  could be  three free $N=1$ supermultiplets, e.g. the supersymmetric sigma model with target space $T^3$.
Here we shall focus on the particular case in which the SCFT $\cal{N}$
is the SCFT on $S^1$ with $c=3/2$ tensored  with a $c=3$ model of six free fermions, which can be thought of as the fermionization of the $N=1$ SCFT on $T^2$.\footnote{Note that these six fermions generate an $su(2) \times su(2)$ Ka\v c-Moody algebra, which allows one to interpret this sigma model as $S^3 \times S^3$.  In this sense the full theory can also be thought of as $AdS_3 \times S^3 \times S^3 \times S^1$.}
 The number of  fermionic fields  will play a  significant role later.

{The constraints from requiring unitarity,  normalizability of physical states, and symmetry under spectral flow
imply that short string physical states now exist only for \cite{gaberdiel-ads} 
\beq
1/2<j<(k+1)/2.
\eeq{sunitarity}
For $k\leq 1$ this excludes many short string states, including the vacuum and graviton with $j=1$.}  
As in the bosonic string, for $k=1, w=1$, there is a  stringy tower of massless states \cite{Gaberdiel:2017oqg},\cite{gaberdiel-ads}.
The results again are in agreement with the effective description of Seiberg and Witten
\cite{Seiberg:1999xz} for the supersymmetric case.

The current algebra is a supersymmetrization of the  affine  $ {SL}(2,\mathbb{R})$ algebra.
Shifting the bosonic currents by fermion bilinears gives a direct product of two commuting sub-algebras,
a free fermion algebra with generators $\psi^\pm,\psi^3$
and an affine  $ {SL}(2,\mathbb{R})$ algebra of level $\kappa =k+2$ with generators $j^\pm_n,j^3_n$.
The nonzero commutation and anticommutation relations are
 \bea
 [j^3_m , j^3_n]&=& -{k+2 \over 2} m \delta_{n+m,0}, \qquad [j^3_m ,  j^\pm_n]= \pm j^\pm_{m+n}, \qquad 
 [j^+_m , j^-_n]= (k+2) m\delta_{m+n,0} -2j^3_{m+n}, \nonumber \\
 \{ \psi^+_r , \psi^-_s\} &=& k\delta_{r+s,0},\qquad \{ \psi^3_r ,\psi^3_s\}= -{k\over 2}\delta_{r+s,0}.
 \eea{m1d}
 The central charge (\ref{ssdfa}) can be written in a form similar to (\ref{sdfa}) as
 \beq
 c_k =
 \frac{3\, \kappa}{\kappa-2}+  \frac{3}{ 2}
 \eeq{therte}
and $k=1$ corresponds to $\kappa=3$.

The quantum states fit into representations of the bosonic $ {SL}(2,\mathbb{R})$ algebra of level $\kappa =k+2$ and of the free fermion algebra, where the fermions can have Ramond or Neveu-Schwarz boundary conditions. 
The spectrum has been analyzed in   \cite{ gaberdiel-ads, Giribet:2007wp, Giveon:1998ns,Pakman:2003cu,Israel:2003ry,Raju:2007uj,Argurio:2000tb}.
We will now summarize the results for the 
 spectrum of one-particle states  for the superstring in AdS$_3$ that we need here, 
 following~\cite{gaberdiel-ads}. 

The physical-state condition in the NS sector with $w$ units of spectral flow is 
\beq 
\tilde{L}_0 -1/2 =0 , \qquad
\tilde{L}_0 = L_0 -w\tilde{J}_0 +kw^2/4. 
\eeq{m1}
The target space energy 
  $E$ and spin $s$ are given by
\beq
{E+s\over 2}=\tilde{J}_0, \qquad  {E-s \over 2}=\tilde{\bar{J}}_0,
\eeq{m1a}
 while 
 \beq
 L_0= -j(j-1)/k + N + h
 \eeq{m1b} 
 where   $N$ is the level of the state in the $SL(2,\mathbb{R})$ current algebra 
 and is given by 
 \beq
 N= {1\over k}\left[\sum_{m=1}^\infty ( j^+_{-m} j^-_m + j^-_{-m} j^+_m -2 j^3_{-m} j^3_m)  + 
 \sum_{r>0} r( \psi^+_{-r} \psi^-_r 
 +\psi^-_{-r} \psi^+_r -2\psi^3_{-r} \psi^3_r)\right] .
 \eeq{m1c}
 The spin of the continuous representation is 
 $j=1/2 +i \sigma$, $\sigma\in \mathbb{R}$, $\sigma>0$.  
 Here $h$ is the (worldsheet) 
 conformal weight for 
   the  internal SCFT ${\cal N}$, which for the free field theory we are considering here is 
$h=N_{\cal N} +h_0$
where   $N_{\cal N}$ is the excitation number for the internal SCFT.
All excitation number are either integer or half integer. 

The GSO projection in
 the NS sector is 
\beq
N+N_{\cal N} +(w+1)/2 \in \mathbb{N}.
\eeq{m1e}
To understand the shift $(w+1)/2$, we 
recall 
that the  three fermionic 
coordinates $\psi^\pm, \psi^3$ transform under spectral flow according to 
\beq
\psi^\pm_r \rightarrow \tilde{\psi}^\pm_r=\psi^\pm_{r\pm w}, \qquad \psi^3_r\rightarrow \tilde{\psi}^3_r=\psi^3_r.
\eeq{m1f}
Consider next the fermionic number operator in the NS sector, 
\beq
N_F\equiv \sum_{r\in \mathbb{N}+1/2} r( \psi^+_{-r} \psi^-_r +\psi^-_{-r} \psi^+_r -2\psi^3_{-r} \psi^3_r).
\eeq{ferm-num} 
Under spectral flow it transforms to
\beq
\tilde{N}_F\equiv \sum_{r\in \mathbb{N}+1/2} r( \tilde\psi^+_{-r} \tilde\psi^-_r 
 +\tilde\psi^-_{-r} \tilde\psi^+_r -2\tilde \psi^3_{-r} \tilde \psi^3_r)=
 \sum_{r\in\mathbb{N}+1/2} r( \psi^+_{-r+w} \psi^-_{r -w}
 +\psi^-_{-r-w} \psi^+_{r+w} -2\psi^3_{-r} \psi^3_r).
 \eeq{m1g}
Some of the fermionic products in this equation are not normal ordered and the shift in (\ref{m1e}) comes from reordering
them. To be concrete, consider the flow with $w=1$. Then 
\beq
\tilde{N}_F= {1\over 2}(\psi^+_{+1/2}\psi^-_{-1/2} - \psi^-_{-1/2}\psi^+_{+1/2}) +
N_F
\eeq{m1h}
We follow ref.~\cite{gaberdiel-ads} and define the  GSO projection in the same way in all flowed sectors {\em in terms of
the flowed fermionic oscillators}: $\tilde{N}_F +N_{\cal N}\in \mathbb{N} +1/2$,  so the shift $(w+1)/2$ 
follows trivially from eq.~(\ref{m1h}).

 The R sector physical-state condition is 
 \beq 
\tilde{L}_0  = L_0 -w\tilde{J}_0 +kw^2/4 =0,
\eeq{m2}
while the GSO projection does not eliminate any conformal weight from the spectrum because of the existence of fermionic
zero-modes. Also, all the excitation numbers in the R sector are always  integers, so that $N+N_{\cal N} \in \mathbb{N}$. 

\vskip .1in

{\em Note that the vacuum energy has been normalized so that the conformal weights of the ground states of both the 
NS and R sector are zero.}

\vskip .1in

The weight $C=-j(j-1)=\sigma ^2 + 1/4$ is the same for both the left and right sectors of the superstring, so 
the normalization of the ground state energy tells us that combining the left and right sectors we obtain integer spin 
$s=\tilde{J}_0- \tilde{\bar{J}}_0$ from the NS-NS and R-R sectors and half-integer spins from the NS-R and R-NS sectors, because for our SCFT
 we can check explicitly that $h_0 -\bar{h}_0 \in \mathbb{Z}$. Of course we would have arrived at the same
result by noticing that the the ground state of the R sector is in the spinorial representation of $SL(2,\mathbb{R})$.

Notice that a consistent GSO projection requires 10 world-sheet fermions (eight plus two that cancel the world-sheet 
supersymmetry ghosts). This is necessary because otherwise the one-loop string integral could not be interpreted as a vacuum energy. 
Technically, the reason is that we need theta-functions with characteristics to appear as $\theta^{4 \mod 8}$. 
If we had chosen ${\cal N}$ to be the SCFT on $T^3$, we would
have only six fermions:  three  from the supersymmetric $SL(2,\mathbb{R})$, three from  $T^3$. Instead we take ${\cal N}$ to be the SCFT on $S^1$
  (with one fermion)
 tensored with the SCFT with 
$6 $ free fermions, which together with $\psi^\pm,\psi^3$
gives 10 fermions in total, as required.

\section{Holographic dual}

Superstring theory in AdS$_3$ is expected to have a 2-d CFT dual. While the supersymmetric case with RR flux has been much discussed, the case with NS-NS flux has been studied less.
In \cite{Argurio:2000tb}, it was conjectured that the dual CFT for the 
superstrings on AdS$_3 \times \cal{N}$ should be a SCFT 
in the moduli space of the
symmetric product ${ \cal{M}}^N/S_N$.  The integer $N$ is associated with
the number of long fundamental strings required to construct the AdS$_3$ 
background and parameterizes the string coupling constant, and $ \cal{M}$ is the spacetime SCFT with $c=6k$ corresponding to   transverse
excitations of a single long string.  There have been many attempts in the literature to identify pieces of this exact SCFT,  some of which suggest a boundary theory that describes the BPS part of the
the spectrum for larger values of $k$.
Clearly, a key issue is  identifying the correct $ \cal{M}$.

To avoid confusion, note that there are four  2-d CFTs which play a role here.
The first is the worldsheet  CFT on AdS$_3 \times \cal{N}$ with critical central charge of $26$ or $15$.
 The second is the worldsheet theory describing the high energy transverse excitations of one long string; an effective theory at large $k$ is given by Liouville theory for the bosonic string and 
super-Liouville theory for the superstring. The third is a class of CFTs capturing only the BPS part of the spectrum. The fourth is the holographic dual CFT, the exact description of the space-time theory at the transition point.

We now  attempt  to identify  the  holographic dual for superstrings with $k=k_c = 1$. 
First, note that at this point  the cosmological constant is of the order of the string scale (though not necessarily of the order of the Plank scale), implying that there are large string corrections.
We may expect to be in the familiar territory  
of a system 
which is rather simple in CFT language but which deviates strongly from the semi-classical geometric picture.

This of course occurs for duals of  AdS$_5$ as well as for the candidates for the duals for higher spin theories on AdS$_4$, the so-called Vasiliev models.
The difference is that  in our case we can compare the exact spectrum of our proposed CFT to the exact bulk spectrum using the results of \cite{mo}, at least in the limit of large $N$.  

The excitations of a long string are described by an effective
CFT, which for large $k$ is a (super-)Liouville theory, that has a continuous spectrum above a gap. For 
$k \leq k_c$ long strings become weakly 
coupled close to the boundary \cite{gkrs}. Since the strings are weakly coupled, we may hope that this description applies at 
$k = k_c$.  For the bosonic string,
the  central charge $1 + 3 Q^2$ of the Liouville theory tends to $c=1$ as $k$ approaches $k_c=3$ (\ref{sdgjjo}),  and the gap 
(\ref{liou}) vanishes so that the spectrum is bounded from below by zero.
For the superstring, the situation is similar.   {The SCFT governing the long string, which for large $k$ is effectively 
given by a super-Liouville theory, is expected to  {exist at any $k$ and} have a
central charge tending to $c=3/2$ as $k$ 
approaches $k_c=1$, with the spectrum again bounded from below by zero.}

Based on these hints, we propose a dual boundary $ SCFT $ theory for the $k=1$  point of the supersymmetric theory. 
The excitations of a long string at $k=1$ are expected to be described by a
$c=3/2$   {SCFT}
 which has a continuous spectrum bounded from below by zero.  The simplest possible theory with these properties is a  free boson plus a free fermion. Therefore
 \emph{we propose that the $ d=2 $ $SCFT $ on the symmetric orbifold   
 $(\mathbb{R}  \times {\cal N})^N/S_N$ 
 is the holographic dual to string theory on $AdS_3 \times {\cal N}$
 at $k=1$.}  
 If ${\cal N}=T^3$ for 
the critical $k=1$ theory, then the dual would be the SCFT on
 $(\mathbb{R}  \times T^3)^N/S_N$, but here we will focus on the case in which ${\cal N}$
is the $S^1$ SCFT tensored with the free fermion theory.
  
 As we will see in the next section, {in the large $N $ limit,  the spectrum of
  the SCFT on
 $(\mathbb{R}  \times T^3)^N/S_N$
   coincides  \emph{precisely}  with  that of  the long strings of the  $ {SL}(2,\mathbb{R})$ WZW model found in \cite{mo}.  This provides non-trivial evidence for the correctness of our duality conjecture. However, there  are  caveats. It could be that the $c=3/2$ SCFT that arises in the limit is not  the free SCFT on $\mathbb{R}$, but a different theory with the same spectrum (or nearly the same spectrum).}\footnote{ 
For the bosonic case, the corresponding limit would give a $c=1$ CFT. One candidate is a free boson, another is the theory discussed by Runkel and Watts in \cite{RW, Sch, HMW}.  According to \cite{RW}, this theory also has a continuous spectrum bounded from below by zero, but missing a discrete set of points.} {{Another possibility is that $\mathbb{R}$ could be replaced by  $\mathbb{R}/\mathbb{Z}_2$.  } This theory has additional normalizable states in addition to those that   arise for $\mathbb{R}$.} Careful consideration of the interactions is needed to determine which $c=3/2$ SCFT is correct here, and what  role the remaining short string states play. We leave this investigation to future work.

The presence  of the $\mathbb{R}$ factor in the boundary CFT  suggests considering a similar factor in the bulk theory. 
Indeed, we shall show in the next section  that the parameter $\sigma$ in the bulk spectrum should be identified  
with $p/\sqrt {2}$ where $p$ is the momentum  in the $\mathbb{R}$ factor in the boundary CFT. 
  Note that if the target space of
the seed theory in our CFT orbifold is 
$\mathbb{R}\times \mathcal{N}$, the momentum $p$ ranges over \draftnote{  the entire real line}.  On the other hand if it is $(\mathbb{R}/\mathbb{Z}_2)\times \mathcal{N}$, there is only one state for each $|p|$ (and there are additional discrete states in the twisted sector).  Certain observables are insensitive to this distinction (c.f.~Section \ref{inter}).

This system of higher-spin massless states may enjoy the conformal invariance that might be expected at a phase transition.
However, there remain massive states in the spectrum at this point (e.g. from continuous representations with $w\ne 1$, or from states with both $N$ and $\bar N$ non-zero), so the full theory is presumably not conformally invariant, {but may have spontaneously broken conformal symmetry.}

\subsection{The permutation orbifold $(\mathbb{R} \times  {\cal N})^N/S_N$.}

The Hilbert space of a permutation orbifold sigma model  $M^N/S_N$ decomposes into twisted sectors labeled by the conjugacy classes of
$S_N$ (see e.g.~\cite{dmvv}). The sigma-model fields are maps $S^1 \times \mathbb{R}\rightarrow M^N/S_N$, and
the $S^1$ coordinate $\sigma$ has periodicity $\pi$. The conjugacy classes are associated with partitions of $N$ into integers: 
$N=\sum_{n=1}^N n M_n$. Each class $[g]$ consists of all permutations of the form
\beq
[g]=(1)^{M_1} (2)^{M_2}.... (N)^{M_N} = \prod_{n=1}^{N} (n)^{M_{n}}.
\eeq{m3}
In the above we denote a cyclic permutation of $n$ elements by $(n)$, with $M_n$ its multiplicity (for instance, a conjugacy 
class of $S_9$ is $(1)^2(2)^2(3)$, with $M_{1} = 2, M_{2} = 2, M_{3}=1$, and all other $M_{i} = 0$). 
Each conjugacy class gives rise to a twisted sector of the symmetric product, where the coordinate $X_i$ in the $i$-th
copy of $M$ obeys the boundary condition
\beq
X_i(\sigma +\pi)= X_{[g](i)} (\sigma).
\eeq{m4}
The untwisted sector is given by $[g]=(1)^N$, for which $[g](i)=i$  $\, \, \forall i$. 
The Hilbert space obtained by quantizing the fields $X_i$ with boundary condition~(\ref{m4}) must be projected to the 
invariant subspace of the commutant $C_g$ of $[g]$~\cite{dmvv}
\beq
C_g=S_{M_1}\times (S_{M_2} \rtimes \mathbb{Z}_2)\times ..... (S_{M_N} \rtimes \mathbb{Z}_N).
\eeq{m5}
Explicitly, the Hilbert  space of the twisted sector $[g]$ is
\beq
\mathcal{H}^{C_g}_g = \otimes_{n=1}^N S^{M_n} \mathcal{H}_{(n)}^{\mathbb{Z}_n}.
\eeq{m6}
The various symbols appearing here are defined as in~\cite{dmvv}: 
\begin{enumerate}
\item $S^k (\mathcal{H})=(\mathcal{H} \otimes..... \mathcal{H})^{Sym}$ is the $k$-fold product of the vector space 
$\mathcal{H}$ symmetrized according to the grading of $\mathcal{H}$. 
\item $\mathcal{H}_{(n)}^{\mathbb{Z}_n}$ is the $\mathbb{Z}_n$-invariant sector of $\mathcal{H}_{(n)}$.
\item $\mathcal{H}_{(n)}$ is the Hilbert space of the sigma model $M^n$ with boundary condition
\beq
X_i(\sigma +\pi)=X_{i+1 \, \text{mod} \, {n}}(\sigma) .
\eeq{m7}
\end{enumerate}
The full Hilbert space of the theory is the direct sum of  (\ref{m6}) over all conjugacy classes.
{Our proposal is that the single-particle states of the bulk with winding $w$ correspond to states where $M_1=N-w$ and $M_w=1$ (and all other $M_n=0$).  Multi-particle states are those with more than one sector $M_{n > 1} \neq 0$.  Notice that at finite $N$, the maximum possible winding $w$ is $w=N$.  As there is no limit on $w$ in the bulk according to classical string theory, it is only in the $N \to \infty$ limit that  our boundary theory could reproduce the bulk spectrum of long strings found in \cite{mo}.  This is as expected, since the limit   $N \to \infty$ should correspond to classical physics in the bulk.}

The sector (\ref{m7}) can be thought of as a map from an $S^1$ with periodic coordinate $\sigma \sim \sigma + n\pi$ to $M$.
Therefore all energies are divided by $n$ and the vacuum energy is also shifted to ${c\over 24}(n-1/n)$ 
(see e.g.\ \cite{af,gaberdiel-symm-orb}). Then, a state with right conformal weight $h$ and left conformal weight $\bar{h}$
 in the conformal sigma model on $M$ 
maps to a state in $\mathcal{H}_{(n)}$ with conformal weights
\beq
h_n = {h \over n} + {c \over 24} \left( n -{1\over n} \right), \qquad 
\bar{h}_n = {\bar{h} \over n} + {c \over 24} \left( n -{1\over n} \right).
\eeq{m8}
In our case $M=\mathbb{R}\times {\cal N}= \mathbb{R}\times S^1\times {\cal F}$ where 
${\cal F}$ is the SCFT of 6 free fermions  
and $c=6$. To find the specific form of $h,\bar{h}$, we can write the conformal weight as a sum of excitation numbers $N+N_{\cal N}$ 
(with $N_{\cal N}=N_{S^1} + N_{\cal F}$) plus 
{the}
 primary weight $h_0$. The excitation numbers
 are each the sum of
a bosonic excitation number, which is always integer, plus a fermionic excitation number, that can be either integer or 
half-integer. We will need only the explicit form of the excitation number in 
the SCFT on
$\mathbb{R}$,
 consisting of a free boson with modes $a_n$ and a free fermion with modes $\psi_m$
\beq
N= \sum_{m=1}^\infty (a_{-m}a_m + m\psi_{-m}\psi_m), \qquad [a_m , a_{n}]=m\delta_{m+n,0},\qquad \{\psi_m , \psi_n\} =
\delta_{m+n,0}.
\eeq{m8a}
The primary weights $h_0,\bar{h}_0 $ are
\beq
h_0={1\over 2} p^2 + h_0^{{\cal N}} , \qquad \bar{h}_0={1\over 2} p^2 + \bar{h}_0^{{\cal N}} 
\eeq{m9}
where $p$ is the momentum in the $\mathbb{R}$ factor.

\subsection{Fermionic boundary conditions}

The sigma model on $M$ is defined only once we specify its spectrum. To do so, we have to decide whether 
fermions have periodic (R) boundary conditions, antiperiodic (NS) boundary conditions, or both. In the first case the Hilbert 
space of the sigma model is only the R sector, in the second it is only the NS sector while in the third it is a sum of R plus NS.
Likewise, we can decide to project out states with odd fermion number in either sector. We want to reproduce the spectrum of 
superstrings on 
AdS$_{3}\times {\cal N}$ with $ {\cal N}=S^1 \times  {\cal F}$, where fermions can be both NS and R and in the sector with 
$w$ units of flow
we project over states with fermion parity $(-1)^F= (-1)^{1+w}$ (see eq.~(\ref{m1e})).
To mimic the string construction as closely as possible, we {include both periodic and antiperiodic fermions in the CFT make a projection on fermion number {\em before} the symmetrization $S_N$. By considering both NS and R sectors and imposing the projection before the symmetrization, we obtain a model that describes transverse oscillations of long strings on AdS$_{3}\times S_1\times {\cal F}$ 
instead of a generic  symmetric-orbifold CFT.
 In Section \ref{equiv} we relate this to an equivalent construction where all fermions have NS boundary conditions.}

Since the fermion parity of the $i$-th copy in $M^N$ and $(-1)^F_i$ and $S_N$ do not commute, we have to carefully examine 
 the possible boundary conditions of fermions. We will do so in the next subsection.

\section{Comparison of spectra}

The spectrum of conformal dimensions given by eq.~(\ref{m1}) at $k=1$ can be written in a form valid for both the NS and
R sectors. In the sector with $w$ units of flow it is
\bea
h &=& {E+s \over 2} = {1\over w} \left( \sigma ^2 +N+N_{\cal N} +h_0 + {a\over 2}\right) + 
{1\over 4} \left( w-{1\over w}\right) , 
\nonumber  \\
\bar{h} &=& {E-s \over 2} = {1\over w} \left( \sigma ^2 +\bar{N}+\bar{N}_{\cal N} +\bar{h}_0 +{b\over 2}\right) + 
{1\over 4} \left( w-{1\over w}\right) .
\eea{m14}
The constants $a,b$ are zero in the NS sector and one in the R sector.
This is the spectrum of single-particle states. We want to identify it with the spectrum of states in 
$\mathcal{H}_{(w)}^{\mathbb{Z}_w}$, because single-particle states map to states where only one copy of the Hilbert space of $M$ in (\ref{m6}) is not in its ground state. 
{Using (\ref{m8}), the definition of the conformal weight given below it, and 
(\ref{m9}) we find, on changing $n\to w$,}
\bea
h_w &=&  {1\over w} \left( {1\over 2} p^2 +N+N_{\cal N} +h_0^{{\cal N}}   \right) + 
{1\over 4} \left( w-{1\over w}\right) , 
\nonumber  \\
\bar{h}_w &=&  {1\over w} \left( {1\over 2}p^2 +\bar{N}+\bar{N}_{\cal N} +\bar{h}_0^{{\cal N}}  
 \right) + {1\over 4} \left( w-{1\over w}\right) .
\eea{m15}
Eqs.~(\ref{m14}) and~(\ref{m15}) are very similar. To prove that they are in fact identical we have to take care of a few
details.

\subsection{Multiplicities}

The excitation number operators $N_{\cal N} $ in~(\ref{m14},\ref{m15})  are evidently the same, as are the barred 
excitation numbers  $\bar N_{\cal N} $. The operator $N$ in~(\ref{m14}) is given in 
eq.~(\ref{m1c}) while the operator $N$ in (\ref{m15}) is given in~(\ref{m8a}). Despite appearances they give the same 
spectrum with the same multiplicities thanks to the no-ghost theorem of~\cite{mo,Pakman:2003cu}. The theorem says that up to null states, 
the physical states of the fermionic string on AdS$_{3}$ obey $J^3_n |Phys\rangle =0$, $\psi^3_n|Phys\rangle =0$ for all $n>0$.
If we denote by $L_n^3, G^3_r$ the Virasoro and world-sheet supersymmetry generators of the sigma model
associated to $J^3$ in $SL(2,\mathbb{R})$ then $\hat{L}_n\equiv L_n-L^3_n$ and $\hat{G}_r\equiv G_r -G^3_r$ commute with 
$J^3_n$, so that every  physical state is a null state plus a linear combination of vectors
\beq
\hat{L}_{n_1}... \hat{L}_{n_I} \hat{G}_{r_1}... \hat{G}_{r_J} | j,m,\alpha\rangle  ,
\eeq{m16}
where the  state $| j,m,\alpha\rangle$ denotes a primary of the current algebra of $SL(2,\mathbb{R})$. The Hilbert space 
generated by these vectors is isomorphic to the Fock space of one fermionic oscillator plus one bosonic oscillator. The natural
isomorphism of basis vectors maps the operator $N$ given in~(\ref{m1c}) into the $N$ given in~(\ref{m8a}).

\subsubsection*{Invariant states of the superstring on AdS$_{3}\times {\cal N}$}

Not all values of $N$, $N_{\cal N} $ are allowed. The single-string states in AdS$_{3} \times {\cal N}$ are 
constrained by the condition
$h-\bar{h} \in \mathbb{Z} +a'/2$, with $a'=0$ in the NS-NS and R-R sectors and $a'=1$ in the NS-R and R-NS sectors. 
Thanks to eq.~(\ref{m14}), this condition becomes, in all sectors,
\beq
N+N_{\cal N} +h_0+ {a\over 2}  - \bar{N}  - \bar{N}_{\cal N} -\bar{h}_0 -{b\over 2} = 
w \left( m +{a-b\over 2}\right) , \qquad m\in \mathbb{Z}.
\eeq{m17}
As we mentioned earlier, the GSO projection imposes the further constraint 
\beq
(-1)^F=(-1)^{w+1}, \qquad (-1)^{\bar{F}}=(-1)^{w+1}.
\eeq{m17a}

\subsection{Invariant states of the symmetric orbifold}
Projecting over invariant states is straightforward for bosonic coordinates but less so for fermions.
Fermions can be twisted in two ways: by the permutation and by the fermionic parity. Consider the cyclic permutation 
$(n)$ inside $[g]=(1)^{M_1}(2)^{M_2} .... (N)^{M_N}$. Up to trivial field redefinitions the boundary conditions for the $n$ 
cyclically-permuted fermions are 
\beq
R: \; \psi_j(\sigma + \pi) =\psi_{j+1} (\sigma) \mod n, \qquad  NS: \; \psi_j(\sigma+\pi) =(-1)^{\delta_{nj }}\psi_{j+1} (\sigma) 
\mod n .
\eeq{m18}
Call $g$ the generator of $\mathbb{Z}_n$, the cyclic permutation of $n$ elements\footnote{The winding number $w$ of the bulk theory corresponds to the cycle length $n$ in the holographic dual discussed in section 4. Hereafter we will write $n$.}; $(-1)^{F_i}$ are the generator of the 
fermionic parities $\mathbb{Z}_2$. Then the R boundary conditions 
in eq.~(\ref{m18}) is a twist by the element $g $ in $\mathbb{Z}_n \rtimes (\mathbb{Z}_2)^n $ while the NS boundary 
condition is a twist by the element $g (-1)^{F_n}$. {\em To find invariant states we must project over the commutant of the 
twist element.} For R boundary condition the projection is obviously
\beq
P= {1\over 2}\left[1+ (-1)^{F}\right] {1\over n} \left(\sum_{j=1}^n g^j \right), \qquad F\equiv \sum_{j=1}^n F_j
\eeq{m19}

For NS, the commutant is found as follows. A product of fermion parities, $P_A =\prod_{j\in A}(-1)^{F_j}$, acts on
fermions as $P \psi_j = \eta_j \psi_j P$ with $\eta_j=-1$ if $j\in A$ and $\eta_j=+1$ if $j\notin A$. The action of $g$ is of course
$g\psi_j=\psi_{j+1} g \mod n$. Call $A=g(-1)^{F_n}$; it acts on fermions as $A\psi_j= \eta_j \psi_{j+1}A$, 
$\eta_j=(-1)^{\delta_{jn}}$ . Call $B=g^k P_A$ then
\beq
AB\psi_j = \eta_{j+k} \epsilon_j \psi_{j+k+1} AB ,\qquad BA\psi_j=\eta_j \epsilon_{j+1} \psi_{j+k+1} BA.
\eeq{m20}
It is easy to see that the commutant of $A$ is the subgroup generated by $1$, $(-1)^F$, and the  elements 
$B_k\equiv g^k \prod_{j=n-k-1}^n (-1)^{F_j}$, $k=1,..n-1$. 
Next, expand the fermionic coordinate in oscillators
\beq
\psi_j(\sigma)= \sum_{r \in \mathbb{Z}+1/2} d_r e^{2ir[\sigma +\pi(j-1)]/n}, \qquad j=1,..,n, 
\qquad \psi_{j+n}(\sigma)\equiv \psi_j(\sigma).
\eeq{m21}
On the oscillators $d_r$, $ B_k d_r = \exp[2ir k\pi/n]d_r  B_k$.

There is one last point to consider. The sigma model on $M$ has left- and right-moving fermions. The GSO projection
acts independently on the left and right movers. On the other hand, the twist acts simultaneously on left movers and right 
movers. So, in the NS-NS sector, the commutant is generated by $1$, $(-1)^F$, $(-1)^{\bar{F}}$, and 
$g^k \prod_{j=n-k-1}^n (-1)^{F_j+\bar{F}_j}$, $k=1,...n-1$. 
In the NS-R sector the commutant is generated by $1$, $(-1)^F$, $(-1)^{\bar{F}}$, $g^k \prod_{j=n-k-1}^n (-1)^{F_j}$, 
$k=1,...n-1$; in the R-NS sector, barred and unbarred operators exchange their role. 

Finally, we can use the identity $(-1)^{F_j}g= g(-1)^{F_{j-1}}$ to write the projectors over invariant states as 
\bea
NS-NS: && P= {1\over 4} [1 +(-1)^F][1+ (-1)^{\bar{F}} ] \sum_{j=0}^{n-1} [g(-1)^{F_n+\bar{F}_n}]^j , \nonumber \\
NS-R: && P= {1\over 4} [1 +(-1)^F][1+ (-1)^{\bar{F}} ] \sum_{j=0}^{n-1} [g(-1)^{F_n}]^j  .
\eea{m22}

Invariant states can be written in a more transparent way, valid for all sectors, by using the action of the commutant 
on the oscillators $d_r$, $\bar{d}_r$. The expansion~(\ref{m21}) holds in the R sector too if the index $r\in \mathbb{Z}$. 
The projections~(\ref{m19}) and (\ref{m22})
keep states invariant under three distinct transformations: 
\beq
d_r\rightarrow -d_r, \qquad \bar{d}_r \rightarrow -\bar{d}_r, \qquad  
\left(d_r,\bar{d}_{\bar{r}} \right)\rightarrow \left(e^{2\pi i  r/n} d_r, e^{-2\pi i \bar{r}/n} \bar{d}_{\bar{r}}\right ). 
\eeq{m23}
The first two fix the left and right fermion parity. To find states invariant under the last transformation we must define
its action on the ground states of the fermionic oscillators, 
$|h_0^{\mathcal{N}}\rangle |\bar{h}_0^{\mathcal{N}}\rangle$. The correct definition turns out to be
\beq
B_1 |h_0^{\mathcal{N}}\rangle |\bar{h}_0^{\mathcal{N}}\rangle = 
(-1)^{a+b}e^{2i\pi (h_0^{\mathcal{N}} - \bar{h}_0^{\mathcal{N}})/n} |h_0^{\mathcal{N}}\rangle |\bar{h}_0^{\mathcal{N}}\rangle,
\eeq{m23a}
with $a,b=0$ in the NS sectors and $a,b=1$ in the R sectors. The resulting constraint is
\beq
N+N_{\cal N}  +h_0^{\mathcal{N}} - \bar{N}  - \bar{N}_{\cal N} - \bar{h}^{\mathcal{N}}_0 \in n \left(\mathbb{Z} +{a-b \over 2}
\right).
\eeq{m24}
This equation matches  eq.~(\ref{m17}) perfectly, once the shift in conformal weights described in the next
subsection is taken into account, so it is natural to define the fermion parity
 of the vacuum states in the $(n)$
sector as $(-1)^F|0\rangle= (-1)^{\bar{F}}|0\rangle=(-1)^{1+n}|0\rangle$ to match eq.~(\ref{m17a}).

\subsubsection{Shift in the conformal weight}
There is one last point to clarify before comparing the spectra in eqs.~(\ref{m14}) and ~(\ref{m15}). Let us examine the
unbarred sector, since the story will be the same for the barred sector. 
There seems to be a discrepancy between the conformal weight $\sigma^2 + h_0 +a/2$ in~(\ref{m14}) and the conformal weight 
$p^2/2+h_0^{{\cal N} } $ 
in~(\ref{m15}). In reality there is no discrepancy. In the NS sector the weights agree once we identify $\sigma=p/\sqrt{2}$. 
When comparing R sectors we must 
recall that we defined $\sigma^2+ h_0$ by subtracting the energy of the R vacuum, 
so that the relation between $\sigma^2+h_0$  and $p^2/2+ h_0^{{\cal N} } $ is
\beq
\sigma^2+ h_0= {p^2\over 2} +h_0^{{\cal N} }   - {n_F\over 16},
\eeq{m25}
where $n_F$ is the number of fermions in the SCFT $\mathbb{R} \times {\cal N}$. To match
weights between~(\ref{m14}) and~(\ref{m15}) we need $n_F/16=1/2$ i.e.~$n_F=8$. 
Taking ${\cal N}= S^1\times {\cal F}$, we have one fermion from the $\mathbb{R}$
sigma model, one from $S^1$ and six from the free fermion factor ${\cal F}$, giving a total of $n_F=8$, as required.
If we had chosen ${\cal N}=T^3$, we would have had
 one fermion from the $\mathbb{R}$
sigma model, one from $S^1$, and two from $T^2$, which would be too few; we would have needed to fermionize 
the two coordinates of
$T^2$ to get another four fermions, so that the $T^2 $ sigma model would be replaced by the free fermion model 
${\cal F}$. The same replacement of the $T^2$ SCFT with ${\cal F}$
 was  needed to make sense of the superstring on 
AdS$_{3}\times S^1 \times T^2$ (see the last paragraph of Section~\ref{sss}).\footnote{We can of course bosonize the four 
the fermions to recover a torus, but the radii of the torus must be fixed to  correspond to the  free fermion theory.}

\subsection{An Equivalent Construction}\label{equiv}
The dual CFT can be defined in an equivalent way by choosing the following fermion boundary conditions 
in the cyclic-permutation sector $(n)$~\footnote{We are indebted to M. Gaberdiel, who raised a number of questions regarding the relationship of our construction to that of \cite{Gaberdiel:2018}. In this subsection, we address the concerns he raised and would like to thank him for his input.} 
\beq
\psi_j(\sigma + \pi) =-\psi_{j+1} (\sigma) \mod n .
\eeq{mr1}

Mapping the cylinder coordinates $\tau,\sigma$ into the complex plane coordinate $z$ 
by the exponential mapping $z=\exp(\tau + 2i\sigma)$
these become the ``pure NS'' conditions 
\beq
\psi_j(e^{2i\pi}z)=\psi_{j+1}(z) \mod n
\eeq{mr2}
 usually considered in the literature on
supersymmetric permutation orbifolds beginning with~\cite{lm}. Eq.~(\ref{mr1})  implies that (after a simple field redefinition) the first of eqs.~(\ref{m18}) 
holds for $n$ even while the second holds for $n$ odd.  

The complete spectrum of the bulk theory given in eqs.~(\ref{m14}) is obtained by {\em not} imposing a GSO projection on
the symmetric-orbifold states. The 
only constraint is the projection over $Z_n$ invariant states. The action on bosons is  well known~\cite{dmvv}; to find
the action on fermions it is convenient to change eq.~(\ref{m21})  and expand the eight fermions as

\beq
\psi_j^I(\sigma)= \sum_{r \in \mathbb{Z}+n/2} d_r^I e^{i2r[\sigma+\pi(j-1)]/n+ i\pi (j-1)}, \qquad \forall j, \qquad I=1,..8.
\eeq{mr2a}
An identical expansion holds for $\bar{\psi}^I_j$ in terms of oscillators $\bar{d}^I_r$. The cyclic permutation $g$ that sends
$j$ into $j+1$ now acts on the  oscillators as
\beq
(d_r^I, \bar{d}_{\bar{r}}^I) \rightarrow (e^{2ir\pi/n +i\pi} d_r^I , e^{2i\bar{r}\pi/n +i\pi} \bar{d}_{\bar{r}}^I ),
\eeq{m2b}
while it acts on the fermionic oscillator ground state as
\beq
g |h_0^{\mathcal{N}}\rangle |\bar{h}_0^{\mathcal{N}}\rangle = 
e^{2i\pi (h_0^{\mathcal{N}} - \bar{h}_0^{\mathcal{N}})/n} |h_0^{\mathcal{N}}\rangle |\bar{h}_0^{\mathcal{N}}\rangle.
\eeq{m2ba} 
The projection over invariant states therefore
changes from~(\ref{m24}) to
\beq
N+N_{\cal N} +h_0^{\mathcal{N}} - \bar{N}  - \bar{N}_{\cal N}  -\bar{h}_0^{\mathcal{N}} \in n (\mathbb{Z} +F/2 -\bar{F}/2).
\eeq{mr2c}

\subsubsection{Odd $n$}
Specifically, when the cyclic permutation $(n)$ has odd length, the symmetric-orbifold fermions are in the NS sector (as defined in \eqref{m18}) so that 
there is no shift in the conformal weight and eq.~(\ref{m25})
gives $h_0=h_0^\mathcal{N}$. 
\begin{description}

\item{Target space bosons:}

Symmetric-orbifold states with an even number of fermions are in one-to-one correspondence 
with long string
states with $w=n$ units of flow in the NS sector of the WZW model. The latter are subject to constraint~(\ref{m17}), which
agrees with eq.~(\ref{mr2c}) once we recall that the WZW model fermions have to be projected according to 
eq.~(\ref{m17a}) $(-1)^F=(-1)^{n+1}=1$ and therefore they too have even fermion number. 

\item{Target space fermions:}

Symmetric-orbifold states with an odd number of fermions have $N+N_\mathcal{N}\in \mathbb{N} +1/2$; this matches the 
WZW fermionic states in the R sector, whose spectrum contains $N+N_\mathcal{N} +1/2$ with 
$N+N_\mathcal{N} \in \mathbb{N}$. The shift by $1/2$ also makes eq.~(\ref{m17}) agree with (\ref{mr2c}). 
To complete the matching of WZW states with orbifold states multiplicities must also agree. To
see that they do,  
we recall that the the symmetric-orbifold states corresponding to the R vacuum of the WZW model are

\beq
d_{-1/2}^I |0\rangle , \qquad I=1,..8.
\eeq{mr3}

These eight states are in the vector representation of the $SO(8)$ rotating the fermion indices $I$. The R vacuum states 
of the WZW model are instead in one of the spinor representations of $SO(8)$. The two representations are mapped into 
each other by $SO(8)$ triality.
\end{description}

\subsubsection{Even $n$}

For even cycle length the symmetric-orbifold fermions are in the R sector (as defined in \eqref{m18}), so their fermion number $N+N_\mathcal{N}$ is 
always integer valued. When comparing conformal weights eq.~(\ref{m25}) gives 
\beq
h_0=h_0^\mathcal{N}-1/2. 
\eeq{mr4}
\begin{description}

\item{Target space bosons:}

The spectrum of the WZW NS sector is projected on states with odd fermion number
$(-1)^F=(-1)^{n+1}=-1$ so conformal dimensions and multiplicities match thanks to the shift~(\ref{mr4}), which also
makes eq.~(\ref{m17}) agree with (\ref{mr2c}). Here the $SO(8)$ 
triality works ``in reverse": the 
ground states of the symmetric orbifold, which are in one of the spinor representations of $SO(8)$, are mapped to the 
ground states of the WZW model, which are in the vectorial representation. 

\item{Target space fermions:}

The spectrum of WZW R-sector states is given by eq.~(\ref{m14}) with $a=1$ (or equivalently $b=1$). 
Thanks to the shift~(\ref{mr4}) and to the fermion number operators in  right hand side of eq.~(\ref{mr2c}) we find again an exact match between these states and the states of the symmetric orbifold.
The ground states in both descriptions belong to a spinorial representation of $SO(8)$.
\end{description}

What we have described in this subsection coincides at $p=0$ and zero winding number on $S^1$ with the states 
found in the construction proposed in~\cite{Gaberdiel:2018} when 
$\mathcal{N}$ is interpreted as $S^3\times S^3 \times S^1$ 
as in footnote 5. The equivalence of the
two constructions of $(\mathbb{R}\times \mathcal{N})^N/S_N$  at $p=0$ and zero $S^1$ winding number 
can be seen for the ``seed theory'' of the symmetric orbifold, i.e. for the case $N=1$, also from the fact that the eight 
fermions in $\mathbb{R}\times \mathcal{N}$ enter the partition function of our construction --which has an  R sector and is 
GSO-projected-- in the combination ${1\over 2}(\vartheta_{00}^4/\eta^4 + \vartheta_{01}^4/\eta^4 +\vartheta_{00}^4/\eta^4)$ 
while they enter in the construction of~\cite{Gaberdiel:2018} as 
$\vartheta_{00}^4/\eta^4$. The two expression are identical thanks 
to the famous Jacobi \emph{\ae quatio identica satis abstrusa.}

\section{Observations on Interactions}\label{inter}

The continuous spectrum of the symmetric orbifold has a density of 
single-particle states of the form $[L + \rho(p)]dp/2\pi$, with $L$ an infrared regulator. 
The physical limit is $L\rightarrow \infty$. The corresponding density of states  in the WZW theory is also of the form
$[L + \rho(p)]dp/2\pi$~\cite{mo2}. Without a detailed computation of the two-point function of primary operators in WZW at $k=k_c$
 we cannot tell whether the range of integration in $p$ is over the positive real numbers or  all real 
numbers (this the source of the uncertainty regarding whether the seed theory of the symmetric orbifold should be {a supersymmetic version of the CFT on
$\mathbb{R}$ or  $\mathbb{R}/\mathbb{Z}_2$, or the CFT of Runkel and Watts   \cite{RW, Sch, HMW},  or something else}).
We can only compare the divergent parts of the spectra, which in fact agree. 

Certain physical observables do not depend on either the 
discrete part of the density of states  or the range of $p$. The reason is that long
string states in $AdS_3$ propagate as asymptotically free particles in an infinite radial direction; in other words, they behave as 
particles moving in a potential $V(\rho)$ such that $\lim_{\rho\rightarrow+\infty}V(\rho)=0$.
A toy model of such a system is a particle moving in a  potential with a  wall located at $\rho=-L$. 
At finite $\rho$ in the limit $L\rightarrow\infty$, the finite-time dynamics of a 
localized wave packet is the same as that  for a free theory, where the momentum ranges from $p=-\infty$ to $p=+\infty$. To 
detect the existence of the wall one has to wait a time $O(2L/v)$, where $v$ the speed of the wave packet.  This would remain true even if there are 
normalizable bound states attached to the wall.  We conjecture that the short strings that remain as $k \to k_c$ are of this nature.

To check this conjecture and match the discrete 
part (and answer the question about the range of integration)
we need to 
perform a careful re-evaluation  of the two point functions at $k=k_c$.  These comparisons are particular cases of a more general problem: comparing 
interactions between the bulk theory and the boundary CFT. We plan to 
do that in  detail in a forthcoming paper; here we will only mention some obvious problems arising from the
attempt to match bulk string amplitudes to boundary CFT correlators.

To simplify the discussion, consider first the interactions of the bosonic string\footnote{Amplitudes of the supersymmetric theory are obtained from those of the bosonic theory by implementing three changes: first, one takes into account the shifting $k \to \kappa = k+2$ in the level; second, one dresses the worldsheet operators with the free fermion contributions; third, one includes in the correlators $n-2$ picture changing operators; see \cite{Gaberdiel:2007vu, Dabholkar:2007ey}.}.
String amplitudes in AdS$_3$ are given in terms of correlation functions of vertex operators integrated over the inserting points $z_i$ ($i=1,2,...n$):
\bea
\mathcal{A}^n_{p_1,p_2,...p_N} = \int \frac{\prod_{i=1}^{n} d^2z_i}{\text{Vol}(PSL(2,\mathbb{C}))} \Big\langle \prod_{i=1}^n \ :V_{j_i}^{w_i}(x_i;z_i): \ \Big\rangle_{\text{WZW}} , 
\eea{Amplitude}
where $\text{Vol}(PSL(2,\mathbb{C}))$ stands for the volume of the conformal Killing group and where in the left hand side 
we omitted for brevity the labels $w_1,w_2,...w_n$ on which the amplitude also depends. We are also omitting ghost 
contributions and picture labels. The complex variables $(x_i,\bar{x}_i)$ are auxiliary variables -- in some sense conjugate to
 $(m_i,\bar{m}_i)$  -- that serve to organize the $SL(2,\mathbb{R})$ representations. From the boundary theory point of view, 
these variables have a clear interpretation as the coordinates of the 
local dual operator $\mathcal{O}_{h_i}(x_i)$~\cite{deBoer:1998gyt}. Each worldsheet operator $V_{j}^{w}(x;z)$ in 
(\ref{Amplitude}) creates a physical state with quantum numbers $j=1/2+i\sigma$ and $w\geq 0$. 

Problems appear when trying to match the bulk observables (\ref{Amplitude}) with 
expectations for the boundary theory because the selection rules for the scattering amplitudes seem to differ from 
those one expects for the boundary observables. For instance, one can compute three point functions with arbitrary values of 
the winding 
numbers $w_i$ ~\cite{Giribet:2001ft}~\footnote{Superstring amplitudes in AdS$_3$ with winding states have also been 
studied in \cite{Giribet:2007wp, cn}}. If $w_i-\sum_{j\neq i}^3 w_j> 1$ for any $i=1,2,3$ they vanish~\cite{mo3}; more 
generally, WZW correlation functions (\ref{Amplitude}) vanish unless the winding numbers $ w_i$ satisfy the bounds 
$w_i-\sum_{j\neq i}^n w_j\leq n-2$ \cite{mo3}. On the other hand, we can compute the three point function of
twist operators that create cycles of length $w_i$ as a function of position on the boundary using the 
techniques of~\cite{prr}. A result of their 
analysis is that three point functions of operators that create the twisted states $g_1,g_2,g_3 \in S_N$ 
can be nonzero only when there exist three elements of $S_N$, $h_i,h_2,h_3$, such that $\prod_{i=1}^3 h^{-1}_i g_i h_i =1$.
It is not obvious that these two conditions are consistent.
Finally, if the  CFT dual is in fact $(\mathbb{R}\times S^1\times {\cal F})^N/S_N$, there will be a selection rule 
associated with a conserved momentum $p= \sqrt{2}\sigma$ in 
$\mathbb{R}$. 
Conservation of momentum
 should make amplitudes with $\sum_i \sigma_i\neq 0$ vanish, yet (\ref{Amplitude}) does not seem to obey this rule.
 We plan to   {address}
 these puzzles in a separate paper.

\section{Conclusions}\label{conc}
 Among the various candidate phases of gravity is a phase in which the string has no tension.  
 The transition point we address here, instead,  contains two chiral trajectories 
of massless excitations of all spins, but it also contains many more massive states whose mass scale derives from the 
original theory on AdS$_3 \times \cal{N}$.
 While the tensionless phase of gravity is still mysterious, the
$k=k_c$ 
  {theory} 
 can be described very concretely. Its  long string spectrum is completely described by a symmetric orbifold of a free
CFT. We do not understand yet fully if the interactions of the WZW superstring at $k=1$ pinpoint the undeformed orbifold CFT 
as {\em the} holographic dual  to the full theory, but we do have encouraging indications that this may be indeed true. 
In any event, the fact that
the {\em complete} spectrum of the  long strings, not just the BPS sector, can be matched to such a simple CFT is remarkable
and novel.  {Finally, it would be interesting to better understand the long string phase that arises for $k<k_c$ and
which has no graviton. }

\

\noindent {\bf Note added.}
Related work has independently been done in  \cite{Gaberdiel:2018}. We thank the authors of \cite{Gaberdiel:2018} for sharing their draft with us prior to publication.

\section*{Acknowledgements}
We thank G. Gaberdiel, R. Gopakumar, D. Kutasov, J. Maldacena, L. Rastelli and Xi Yin for useful 
discussions at various stages of  development of this paper.
The work of MK and GG is supported in part by the NSF through grant PHY-1214302.  
MK acknowledges membership at the NYU-ECNU Joint Physics Research Institute in
Shanghai. The work of MP is supported in part 
by NSF through grant PHY-1620039. The work of ER was partially supported 
by the Israeli Science Foundation Center of Excellence and by the I Core Program ``The Quantum Universe," sponsored by the Planning and Budgeting Committee and the Israeli Science Foundation. The work of CH is  supported by the EPSRC 
program grant ``New Geometric Structures from String Theory" EP/K034456/1 and the STFC grant ST/L00044X/1.  CH and ER thank
CCPP  for its  hospitality while part of this work was done.

  \end{document}